# Quark star model with charge distributions

Manuel Malaver

Departamento de Ciencias Básicas, Universidad Marítima del Caribe, Catia la Mar, Venezuela

**Email address**

mmf.umc@gmail.com

**Abstract**

In this paper, we studied the behavior of relativistic objects with anisotropic matter distribution in the presence of an electric field considering a gravitational potential Z(x) of Thirukkanesh and Ragel (2013) which depends on an adjustable parameter n. The equation of state presents a quadratic relation between the energy density and the radial pressure. New exact solutions of the Einstein-Maxwell system are generated. A physical analysis of electromagnetic field indicates that is regular in the origin and well behaved. We show as a variation of the adjustable parameter n causes a modification in the charge density, the radial pressure and the mass of the stellar object.

**Keywords**

Relativistic Objects, Electric Field, Gravitational Potential, Adjustable Parameter, Einstein-Maxwell System, Charge Density

## 1. Introduction

One of the fundamental problems in the general theory of relativity is finding exact solutions of the Einstein field equations [1,2]. Some solutions found fundamental applications in astrophysics, cosmology and more recently in the developments inspired by string theory [2]. Different mathematical formulations that allow to solve Einstein´s field equations have been used to describe the behavior of objects submitted to strong gravitational fields known as neutron stars, quasars and white dwarfs [3,4,5].

In the construction of the first theoretical models of relativistic stars are important the works of Schwarzschild [6], Tolman [7], Oppenheimer and Volkoff [8]. Schwarzschild [6] found analytical solutions that allowed describing a star with uniform density, Tolman [7] developed a method to find solutions of static spheres of fluid and Oppenheimer and Volkoff [8] used Tolman's solutions to study the gravitational balance of neutron stars. It is important to mention Chandrasekhar's contributions [9] in the model production of white dwarfs in presence of relativistic effects and the works of Baade and Zwicky [10] who propose the concept of neutron stars and identify an astronomic dense objects known as supernovas.

The physics of ultrahigh densities is not well understood and many of the strange stars studies have been performed within the framework of the MIT bag model [11]. In this model, the strange matter equation of state has a simple linear form given by $p = \frac{1}{3}(\rho - 4B)$ where $\rho$ is the energy density, p is the isotropic pressure and B is the bag constant. However, in theoretical works of realistic stellar models [12-15] it has been suggested that superdense matter may be anisotropic, at least in some density ranges. The existence of anisotropy within a star can be explained by the presence of a solid core, phase transitions, a type III super fluid, a pion condensation [16] or other physical phenomena. In such systems, the radial pressure is different from the tangential pressure. This generalization has been used in the study of the balance and collapse of compact spheres [17-20].

Many researchers have used a great variety of mathematical techniques to try to obtain exact solutions for quark stars within the framework of MIT bag model, since it has been demonstrated by Komathiraj and Maharaj [21], Malaver [22], Thirukkanesh and Maharaj [23] and Thirukkanesh and Ragel [24]. Feroze and Siddiqui [25] and Malaver [26] consider a quadratic equation of state for the matter distribution and specify particular forms for the gravitational potential and electric field intensity. Mafa Takisa and Maharaj [27] obtained new exact solutions to the Einstein-Maxwell system of equations with a polytropic equation of state. Thirukkanesh and Ragel [28] have obtained particular models of anisotropic fluids with polytropic equation of state which are consistent with the reported experimental observations. More recently, Malaver [29,30] generated new exact solutions to the Einstein-Maxwell system considering Van der Waals modified equation of state with and without polytropical exponent. Mak and Harko [31] found a relativistic model of strange quark star with the suppositions of spherical symmetry and conformal Killing vector.

Our objective in this paper is to generate a new class for charged anisotropic matter with an equation of state that presents a quadratic relation between the energy density and the radial pressure in static spherically symmetric spacetime using a gravitational potential $Z(x)$ of Thirukkanesh and Ragel [24] which depends on an adjustable parameter n. We have obtained some new classes of static spherically symmetrical models of charged matter where the variation of the parameter n modifies the radial pressure, charge density and the mass of the compact objects. This article is organized as follows, in Section 2, we present Einstein´s field equations. In Section 3, we make a particular choice of gravitational potential $Z(x)$ that allows solving the field equations and we have obtained new models for charged anisotropic matter. In Section 4, a physical analysis of the new solutions is performed. Finally in Section 5, we conclude.

## 2. Einstein Field Equations

Consider a spherically symmetric four dimensional space time whose line element is given in Schwarzschild coordinates by

$$ds^2 = -e^{2\nu(r)}dt^2 + e^{2\lambda(r)}dr^2 + r^2(d\theta^2 + \sin^2\theta d\varphi^2)  \quad (1)$$

Using the transformations, $x = cr^2$, $Z(x) = e^{-2\lambda(r)}$ and $A^2 y^2(x) = e^{2\nu(r)}$ with arbitrary constants A and c, suggested by Durgapal and Bannerji [32], the Einstein field equations as given in (1) are

$$\frac{1-Z}{x} - 2\dot{Z} = \frac{\rho}{c} + \frac{E^2}{2c} \quad (2)$$

$$4Z\frac{\dot{y}}{y} - \frac{1-Z}{x} = \frac{p_r}{c} - \frac{E^2}{2c} \quad (3)$$

$$4xZ\frac{\ddot{y}}{y} + (4Z + 2x\dot{Z})\frac{\dot{y}}{y} + \dot{Z} = \frac{p_t}{c} + \frac{E^2}{2c} \quad (4)$$

$$\sigma^2 = \frac{4cZ}{x}(x\dot{E}+E)^2 \quad (5)$$

Where $\rho$ is the energy density, $p_r$ is the radial pressure, $E$ is electric field intensity, $\sigma$ is the charge density, $p_t$ is the tangential pressure and dot differentiations with respect to x.

With the transformations of [32], the mass within a radius r of the sphere take the form

$$M(x) = \frac{1}{4c^{3/2}}\int_0^x \sqrt{x}\rho(x)dx \quad (6)$$

In this paper, we assume the following equation of state

$$p_r = \alpha \rho^2 \quad (7)$$

Here $\alpha$ is arbitrary constant.

## 3. The Models

Motivated by Thirukkanesh and Ragel [24], we take the form of the gravitational potential, $Z(x)$ as $Z(x) = (1-ax)^n$ where α is a real constant and n is an adjustable parameter. This potential is regular at the origin and well behaved in the interior of the sphere. For the electric field we make the choice

$$E^2 = \frac{nx}{(1+ax)^2} \quad (8)$$

This electric field is finite at the centre of the star and remains continuous in the interior. In this paper, we have considered the particular cases for n=1, 2, 3.

For the case n=1, using $Z(x)$ and eq. (8) in eq.(2) we obtain

$$\rho = \frac{6ac + (12a^2c-1)x + 6a^3cx^2}{2(1+ax)^2} \quad (9)$$

Substituting (9) in eq.(7), the radial pressure can be written in the form

$$P_r = \alpha \frac{[6ac + (12a^2c-1)x + 6a^3cx^2]^2}{4(1+ax)^4} \quad (10)$$

Using (9) in (6), the expression of the mass function is

$$M(x) = \frac{a^2\sqrt{ax}[4a^4cx^2 + (4a^3-a)x - 3] + 3a^2(1+ax)\arctan\sqrt{ax}}{8c^{3/2}a^4\sqrt{a}(1+ax)} \quad (11)$$

With (8) and $Z(x)$ in (5), the charge density is

$$\sigma^2 = \frac{c(1-ax)(3+ax)^2}{(1+ax)^4} \quad (12)$$

The tangential pressure is given by

$$P_t = 4xc(1-ax)\frac{\ddot{y}}{y} + 2c(2-3ax)\frac{\dot{y}}{y} - ac - \frac{x}{2(1+ax)^2} \quad (13)$$

Substituting (10), (8) and $Z(x)$ in (3), we have

$$\frac{\dot{y}}{y} = \frac{\alpha[6ac + (12a^2c - 1)x + 6a^3cx^2]^2}{16c(1+ax)^4(1-ax)} - \frac{x}{8c(1+ax)^2(1-ax)} + \frac{a}{4(1-ax)} \quad (14)$$

Integrating (14), we obtain

$$y(x) = c_1(1+ax)^A(-1+ax)^B \exp[F(x)] \quad (15)$$

where

$$A = -\frac{576\alpha a^4 c^2 - \alpha + 8a}{256 a^3 c}, \quad B = -\frac{\alpha - 8a + 64 a^3 c}{256 a^3 c} \quad \text{and}$$

$$F(x) = -\left( (24a^3 + 288\alpha a^4 c + 5184\alpha a^6 c^2 + 3\alpha a^2)x^2 \right.$$
$$+ (48a^2 + 144\alpha a^3 c - 3\alpha a + 5184\alpha a^5 c^2)x \quad (16)$$
$$\left. + 2304\alpha a^4 c^2 - 2\alpha + 48\alpha a^2 c + 24a \right) \Big/ 384 ca^3 (1+ax)^3$$

The metric functions $e^{2\lambda}$ and $e^{2\nu}$ can be written as

$$e^{2\lambda} = \frac{1}{1-ax} \quad (17)$$

$$e^{2\nu} = A^2 c_1^2 (1+ax)^{2A} (-1+ax)^{2B} \exp[2F(x)] \quad (18)$$

With n=2, the expression for the energy density is

$$\rho = \frac{6ac + (7a^2c - 1)x - 4a^3cx^2 - 5a^4cx^3}{(1+ax)^2} \quad (19)$$

replacing (19) in (7), we have for the radial pressure

$$P_r = \alpha \frac{[6ac + (7a^2c - 1)x - 4a^3cx^2 - 5a^4cx^3]^2}{4(1+ax)^4} \quad (20)$$

and the mass function is

$$M(x) = \frac{a^2\sqrt{ax}[2a^4cx^2 - 2a^5x^3c + (4a^3c - a)x - 3] + 3a^2(1+ax)\arctan\sqrt{ax}}{4c^{3/2}a^4\sqrt{a}(1+ax)} \quad (21)$$

For the tangential pressure the charge density is given by

$$\sigma^2 = \frac{2c(1-ax)^2(3+ax)^2}{(1+ax)^4} \quad (22)$$

$$P_t = 4xc(1-ax)^2 \frac{\ddot{y}}{y} + 4c(1-ax)(1-2ax)\frac{\dot{y}}{y} - 2ac(1-ax) - \frac{x}{(1+ax)^2} \quad (23)$$

The eq. (3) becomes

$$\frac{\dot{y}}{y} = \frac{\alpha[6ac + (7a^2c - 1)x - 4a^3cx^2 - 5a^4cx^3]^2}{4c(1+ax)^4(1-ax)^2} - \frac{x}{4c(1+ax)^2(1-ax)^2} + \frac{a(2-ax)}{4(1-ax)} \quad (24)$$

Integrating (24), we obtain

$$y(x) = c_2(1+ax)^C(-1+ax)^D \exp[G(x)] \quad (25)$$

where for convenience we have let $C = \frac{5\alpha}{8a}$, $D = -\frac{20\alpha a^2 c + 2a + 5\alpha}{8a}$ and

$$G(x) = \left( 150\alpha a^9 c^2 x^5 + 300\alpha a^8 c^2 x^4 + (30\alpha a^5 c - 6a^6 c - 6\alpha a^7 c^2)x^3 \right.$$
$$+ (24\alpha a^4 c + 3a^3 - 18 a^5 c)x^2 + (6a^2 - 2\alpha a - 18 a^4 c - 42\alpha a^3 c - 168\alpha a^5 c^2)x \quad (26)$$
$$\left. + 3a - 6a^3 c - 6\alpha a^4 c^2 - 36\alpha a^2 c - \alpha \right) \Big/ 24 a^3 c (1+ax)^3 (-1+ax)$$

The metric functions $e^{2\lambda}$ and $e^{2\nu}$ are given by

$$e^{2\lambda} = \frac{1}{(1-ax)^2} \quad (27)$$

$$e^{2\nu} = A^2 c_2^2 (1+ax)^{2C} (-1+ax)^{2D} \exp[2G(x)] \quad (28)$$

With n=3, we can find the following expressions for $\rho, P_r, M(x), \sigma^2, P_t, e^{2\lambda}$ and $e^{2\nu}$

$$\rho = \frac{18ac + (6a^2c - 3)x - 28a^3cx^2 - 2a^4cx^3 + 14a^5cx^4}{2(1+ax)^2} \quad (29)$$

$$P_r = \alpha \frac{[18ac + (6a^2c - 3)x - 28a^3cx^2 - 2a^4cx^3 + 14a^5cx^4]^2}{4(1+ax)^4} \quad (30)$$

$$M(x) = \frac{a^2\sqrt{ax}[4a^6cx^4 - 8a^5x^3c + (12a^3c - 6a)x - 9] + 9a^2(1+ax)\arctan\sqrt{ax}}{8c^{3/2}a^4\sqrt{a}(1+ax)} \quad (31)$$

$$\sigma^2 = \frac{3c(1-ax)^3(3+ax)^2}{(1+ax)^4} \quad (32)$$

$$P_t = 4xc(1-ax)^3 \frac{\ddot{y}}{y} + 4c(1-ax)^2(2-5ax)\frac{\dot{y}}{y} - 3ac(1-ax)^2 - \frac{3x}{2(1+ax)^2} \quad (33)$$

$$e^{2\lambda} = \frac{1}{(1-ax)^3} \quad (34)$$

$$e^{2\nu} = A^2 c_3^2 (1+ax)^{2E} (-1+ax)^{2F} \exp[2H(x)] \quad (35)$$

Again for convenience we have let

$$E = \frac{12a - 3195\alpha c^2 a^4 - \alpha(6a^2c-3)^2 + 238 a^2 \alpha(6a^2c-3)c}{512 \, ca^3}$$

$$F = \frac{-12a - 773\alpha c^2 a^4 + \alpha(6a^2c-3)^2 - 128 \, ca^3 - 238 a^2 \alpha(6a^2c-3)c}{512 \, ca^3}$$

and

$$\begin{aligned}H(x) = -[\,&5400\alpha a^{11}c^2 x^7 - 8280\alpha a^{10}c^2 x^6 - 24480\alpha c^2 a^9 x^5 + \\ &(36a^5 - 4356 a^8 \alpha c^2 + 2178\alpha a^6 c + 192 a^7 c - 108\alpha a^8 + 108\alpha a^6 - 27\alpha a^4 + 22143\alpha c^2 a^8)x^4 + \\ &+ (540\alpha a^7 c^2 - 270\alpha a^5 c - 18\alpha a^2 c + 9\alpha a^3 + 36 a^4 + 288 c a^6 + 32727\alpha c^2 a^7)x^3 \\ &+ (36 a^4 + 8508\alpha a^6 c^2 - 4254\alpha a^4 c + 180\alpha a^6 c^2 - 180\alpha a^4 c + 45\alpha a^2 - 288 c a^5 - 19009\alpha c^2 a^6)x^2 \\ &+ (108 a^2 - 420\alpha a^5 c^2 + 210\alpha a^3 c - 252\alpha a^5 c^2 + 252\alpha a^3 c - 63\alpha a - 672 c a^4 - 13897\alpha c^2 a^5)x \\ &+ 72 a + 5288\alpha^2 c^2 a^4 - 4704\alpha a^4 c^2 + 2352\alpha a^2 c - 144\alpha a^4 c^2 + 144\alpha a^2 c - 36\alpha - 288 c a^3\,] \Big/ 768 c a^3 (1+ax)^3(-1+ax)^2\end{aligned}$$ (36)

## 4. Physical Analysis

In this section we discuss the physical properties that have to be satisfied by the realistic star [28]. With n=1, the gravitational potentials are regular at the origin since $e^{2\nu(0)} = A^2 c_1^2 e^{\frac{2304\alpha a^4 c^2 - 2\alpha + 48\alpha a^2 c + 24 a}{384 c a^3}}$ and $e^{2\lambda(0)} = 1$ are constants and $(e^{2\lambda(r)})' = (e^{2\nu(r)})' = 0$ at r=0. In the center $\rho(0) = 3ac$. $\rho$ is positive if a > 0.

For the case n=2, $e^{2\lambda(0)} = 1$,

$$e^{2\nu(0)} = A^2 c_3^2 e^{\frac{72a + 5288\alpha^2 c^2 a^4 - 4704\alpha a^4 c^2 + 2352\alpha a^2 c - 144\alpha a^4 c^2 + 144\alpha a^2 c - 36\alpha - 288 c a^3}{768 c a^3}}$$

in the origin and $(e^{2\lambda(r)})'_{r=0} = (e^{2\nu(r)})'_{r=0} = 0$. Again the gravitational potential is regular at $r=0$. In the center $\rho(0) = 9ac$.

The following figures represent the graphs of $P_r$, $\sigma$, $M(x)$ and the square of speed of sound $v_{sr}^2$ for some values of $\alpha$, $a$ and C. To maintain of causality, the square of sound speed defined as $v_{sr}^2 = \dfrac{dP_r}{d\rho}$ should be within the limit $0 \le v_{sr}^2 \le 1$ in the interior of the star. In figures 1, 2, 3 and 4 we generated the plots for the radial pressure, radial speed of sound, charge density and mass, respectively with $\alpha = 1/4$, $a = 0.2$ and C=1. We have considered the cases n=1,2,3. The metric for this model is

$$ds^2 = -A^2 y^2(x) dt^2 + \frac{1}{(1-ax)^n} dr^2 + \frac{x}{c}(d\theta^2 + \sin^2\theta d\varphi^2)$$ (37)

$e^{2\nu(0)} = A^2 c_2^2 e^{\frac{3a - 6a^3 c - 6\alpha a^4 c^2 - 36\alpha a^2 c - \alpha}{24 c a^3}}$, in the origin $r=0$, $(e^{2\lambda(r)})'_{r=0} = (e^{2\nu(r)})'_{r=0} = 0$. This shows that the potential gravitational is regular in the origin. In the centre $\rho(0) = 6ac$.

With n=3, $e^{2\lambda(0)} = 1$,

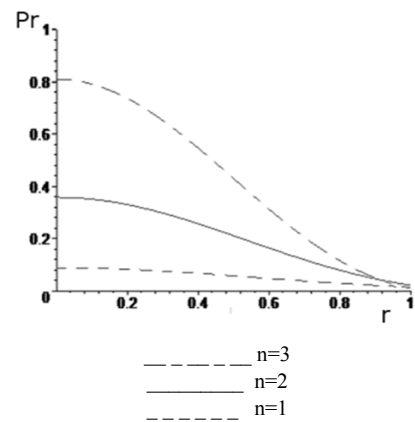

*Fig 1. Radial Pressure.*

In Fig.1, the radial pressure is finite and decreasing for three studied cases. In Fig.2, the condition $0 \le v_{sr}^2 \le 1$ maintained inside the stellar interior. In Fig. 3, that represent charge density, we observe that is continuous, finite and monotonically decreasing function. In Fig.4, the mass function is strictly increasing function, continuos and finite.

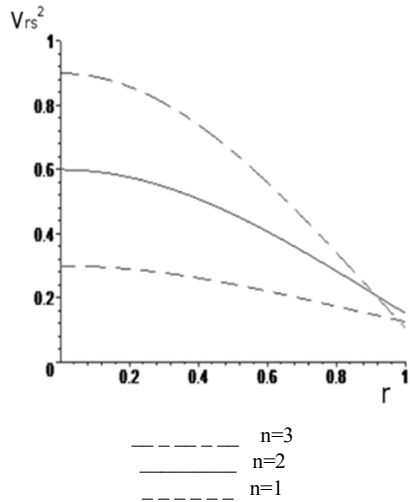

*Fig 2. Radial speed of sound.*

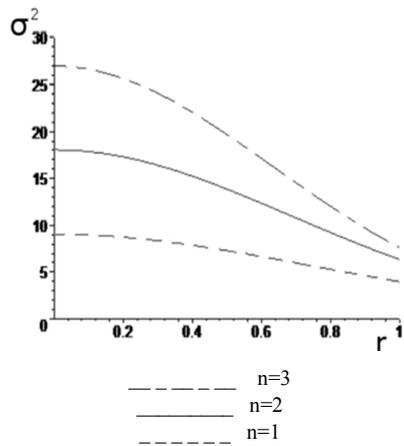

*Fig 3. Charge density.*

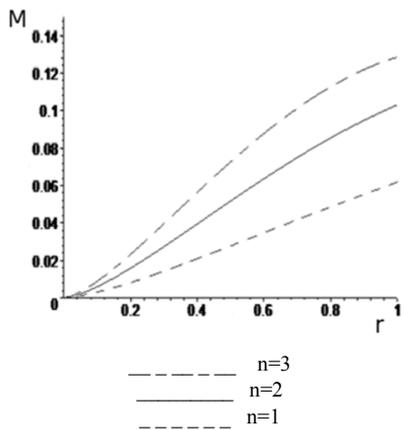

*Fig 4. Mass.*

## 5. Conclusion

In this paper, we have generated new exact solutions to the Einstein-Maxwell system considering a gravitational potential Z what depends on an adjustable parameter n and an equation of state that presents a quadratic relation between the energy density and the radial pressure. The new obtained models may be used to model relativistic stars in different astrophysical scenes. The charged relativistic solutions to the Einstein-Maxwell systems presented are physically reasonable. The charge density σ is nonsingular at the origin and the radial pressure is decreasing with the radial coordinate. The mass function is an increasing function, continuous and finite and the condition $0 \leq v_{sr}^2 \leq 1$ is maintained inside the stellar interior. The gravitational potentials are regular at the centre and well behaved.

We show as a modification of the parameter n of the gravitational potential affects the electrical field, charge density and the mass of the stellar object. The models presented in this article may be useful in the description of relativistic compact objects with charge, strange quark stars and configurations with anisotropic matter.